\documentclass[twocolumn,prl,showpacs]{revtex4}
\usepackage{psfrag}
\usepackage{graphicx}
\usepackage{amsmath}
\usepackage{amssymb}
\usepackage{times}

\begin{document}
\title{Anomalous Conductance Oscillations and Half-Metallicity in
  Atomic Ag-O Chains} \author{Mikkel Strange$^1$,
  Kristian S. Thygesen$^1$, James P. Sethna$^2$, and Karsten
  W. Jacobsen$^1$} \affiliation{$^1$Center for Atomic-scale Materials
  Design, Department of Physics \\
  Technical University of Denmark, DK - 2800 Kongens Lyngby, Denmark\\
  $^2$Laboratory of Atomic and Solid State Physics, Cornell
  University, Ithaca, NY 14853-2501, USA}

\date{\today}

\begin{abstract}
  Using spin density functional theory we study the electronic and
  magnetic properties of atomically thin, suspended chains containing
  silver and oxygen atoms in an alternating sequence. Chains longer
  than 4 atoms develop a half-metallic ground state implying fully spin
  polarized charge carriers.  The conductances of the chains exhibit
  weak even-odd oscillations around an anomalously low value of
  $0.1G_0$ ($G_0 = 2e^2h$) which coincide with the averaged
  experimental conductance in the long chain limit. The unusual
  conductance properties are explained in terms of a resonating-chain
  model which takes the reflection probability and phase-shift of a
  single bulk-chain interface as the only input. The model also
  explains the conductance oscillations for other metallic chains.
\end{abstract}

\maketitle

Atomically thin metallic chains can be formed by pulling a metal
contact apart using e.g. a scanning tunneling microscope or a
mechanically controlled break
junction~\cite{agrait_report,sorensen}. Due to the low coordination
number of the involved atoms, these ultimate one-dimensional quantum
wires exhibit high mechanical stability and are chemically far more
reactive than their bulk
counterparts~\cite{bahn:chain_chemistry,bahn_agrait,oxygen_clamps}.
This, combined with their ability to sustain extremely large current
densities, makes atomic chains interesting from both a fundamental and
a technological point of view.

At sufficiently low temperatures the conduction electrons in a metal
can travel hundreds of nanometers without losing their quantum
mechanical phase~\cite{agrait_report}.  Under such conditions, the
wave nature of electrons becomes important and manifests itself
directly in macroscopic quantities such as the electrical
conductivity.

As first predicted by density functional theory (DFT) calculations,
the conductances of suspended atomic chains of some elements like C,
Na, and Au oscillate with a period of two as the number of atoms, $N$,
in the chain is varied~\cite{lang97,lang98}. This behavior can be
understood from a local charge neutrality condition which implies that
the Fermi level must be aligned with the center of a resonance for odd
$N$ and lie mid between two resonances for even $N$,
respectively. Deviations from these even-odd oscillations naturally
occur for metals with more complex valence configurations such as Al
and Pt~\cite{thygesen_alu,PhysRevB.70.113107}. Common to the
oscillations observed for all the homogeneous metal chains is that the
conductance maximum (per contributing channel) is always close to
$1G_0$ while the oscillation amplitude varies significantly from $\sim
0.05G_0$ in the case of Au~\cite{brandbyge_osc,PhysRevB.70.113107} to
$\sim 0.5G_0$ in the case of Al and C chains
\cite{thygesen_alu,lang_c}. We show here, that this behavior can be
understood in terms of the reflection probability and phase shift at a
single bulk-chain interface.

In a recent experiment Thijssen \emph{et al.} found that the presence
of oxygen greatly enhances chain formation when a silver contact is
broken at low temperature~\cite{thijssen:026806}. It was proposed that
the resulting chains, which can exceed 2 nm in length, consist of Ag
and O atoms repeated in an alternating fashion~\cite{njp:ago},
consistent with previous DFT calculations for Au-O
chains~\cite{bahn:chain_chemistry,oxygen_clamps}.  Interestingly, the
conductance averaged over many chains and plotted as a function of
chain length showed an exponential-like decrease from $1G_0$ to
$0.1G_0$ after which the conductance stayed constant over almost 1
nm. This is indeed a surprising result: the initial decrease in
conductance suggests a chain with a band gap, however, this is not in
line with the saturation at a finite conductance for longer chains. On
the other hand, a length-independent conductance of $0.1G_0$ does not
comply with the results for metallic chains which exhibit conductance
oscillations with a maximum on the order of $1G_0$. We mention that
silver, gold, and nickel contacts with one or two oxygen atoms
adsorbed have previously been
studied.~\cite{ishida,qi_prl,zhang:prl,oxygen_clamps,palacios} In the
following we show that the conductance behavior follows from the
simple phase-shift model.

\begin{figure}[!h]
   \includegraphics[width=0.95\linewidth]{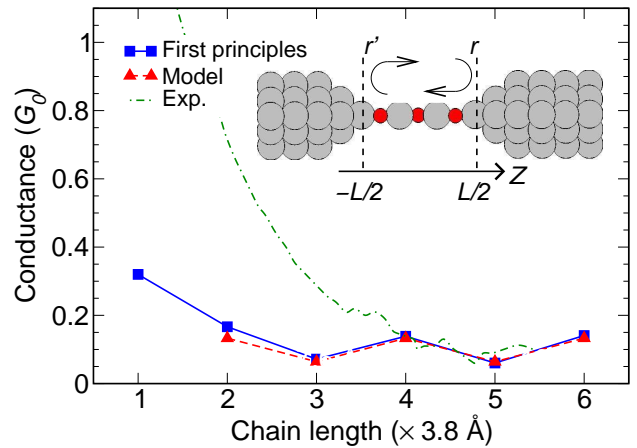}
   \caption{\label{fig.G_N}(Color online.) Calculated conductance as a
     function of chain length (blue squares). The length of the chain
     is given in units of $3.8 \AA$ corresponding to the length of a
     Ag-O unit. Also shown is the result of the resonating-chain model
     (red triangles) and a fit to the experimental data of
     Ref.~\protect\onlinecite{thijssen:026806}. The supercell used to model
     the suspended chains is also shown.}
\end{figure}

In this Letter we present spin DFT calculations showing that
alternating silver-oxygen chains (ASOCs) exhibit small-amplitude
($<0.07G_0$) conductance oscillations around a value of $0.1G_0$ for
long chains. The origin of these anomalous conductance oscillations is
traced to an additional phase picked up by an electron being reflected
at the end of the chain. Furthermore, we find the ASOC to be half-metallic
meaning that only electrons with a given spin direction can pass
through the chain making it a perfect spin-valve.

To describe the chains we use the supercell shown in the inset of
Fig.~\ref{fig.G_N}. It contains the oxygen terminated ASOC suspended
between four-atom silver pyramids, which are attached to (111) silver
surfaces. Periodic boundary conditions are imposed in all
directions. Both the silver pyramid and the silver-oxygen chain have
been relaxed using the Dacapo plane wave DFT code~\cite{dacapo} to
obtain the most stable geometry~\cite{details}.

Assuming phase-coherent transport, the conductance of the chains in
the limit of low-temperature and -bias, is given by the
Landauer-B{\"u}ttiker formula, $G=G_0T(\varepsilon_F)$, where
$T(\varepsilon)$ is the energy-dependent transmission function. The
latter is calculated using the Non-equilibrium Green's function (NEGF)
method described in Refs.~\cite{thygesen_chemphys,benchmark_strange}
where the Kohn-Sham Hamiltonian of the relaxed chains is obtained
from the Siesta code~\cite{details2}.  

In Fig.~\ref{fig.G_N} we show the calculated conductance (blue
squares) of the relaxed oxygen-terminated ASOCs as a function of the
number of oxygen atoms, $N$, in the chain. We note in passing that the
calculated conductances of silver-terminated chains are significantly
lower than the experimental value of $0.1G_0$~\cite{njp:ago}, but
these chains turn out to be less stable than oxygen-terminated chains
and will probably break in the Ag-Ag bond when elongated.  As
indicated on Fig.~\ref{fig.G_N} we have considered relaxed chains with
a length up to $23.0~\text{\AA}$, with the Ag-O bond length lying in
the range $1.95-2.10$~\AA.  For comparison the bond length of the
infinite linear alternating Ag-O wire is found to be $1.95$~\AA.

Fig.~\ref{fig.G_N} also contains a fit to the experimental data
presented in Ref.~\cite{thijssen:026806}. The experimental data is
obtained by averaging the conductance trace of thousands of chains,
and features such as small-amplitude conductance oscillations are not
visible. For long chains, however, there is excellent agreement
between the measured and calculated (average) conductance. The
significantly larger conductances found in the experiments for shorter
chains are presumably due to contributions from pure Ag chains which
have a conductance around $1G_0$.

The calculated conductance of a single oxygen atom, $N=1$, is $0.3G_0$
which is in good agreement with the calculated results in
Ref.~\cite{ishida}.  For $N>1$ the conductance starts to oscillate
with a period of two Ag-O units around an average value of $0.1G_0$.
For $N=1$ we find a non-magnetic ground state, while for $N>1$ the
ferromagnetic state is energetically favored. For large $N$ the
electronic structure of the chain converges toward that of the
infinite alternating Ag-O chain which has a ferromagnetic ground state
with a magnetic moment of $1\mu_B$ per Ag-O unit mainly localized at
oxygen. We find the energy gain with respect to the non-magnetic state
to be $0.12$~eV per Ag-O unit.  This breaking of the spin symmetry is the
reason for the lower conductance found for the chains with $N>1$.

\begin{figure}[!h]
   \includegraphics[width=0.95\linewidth]{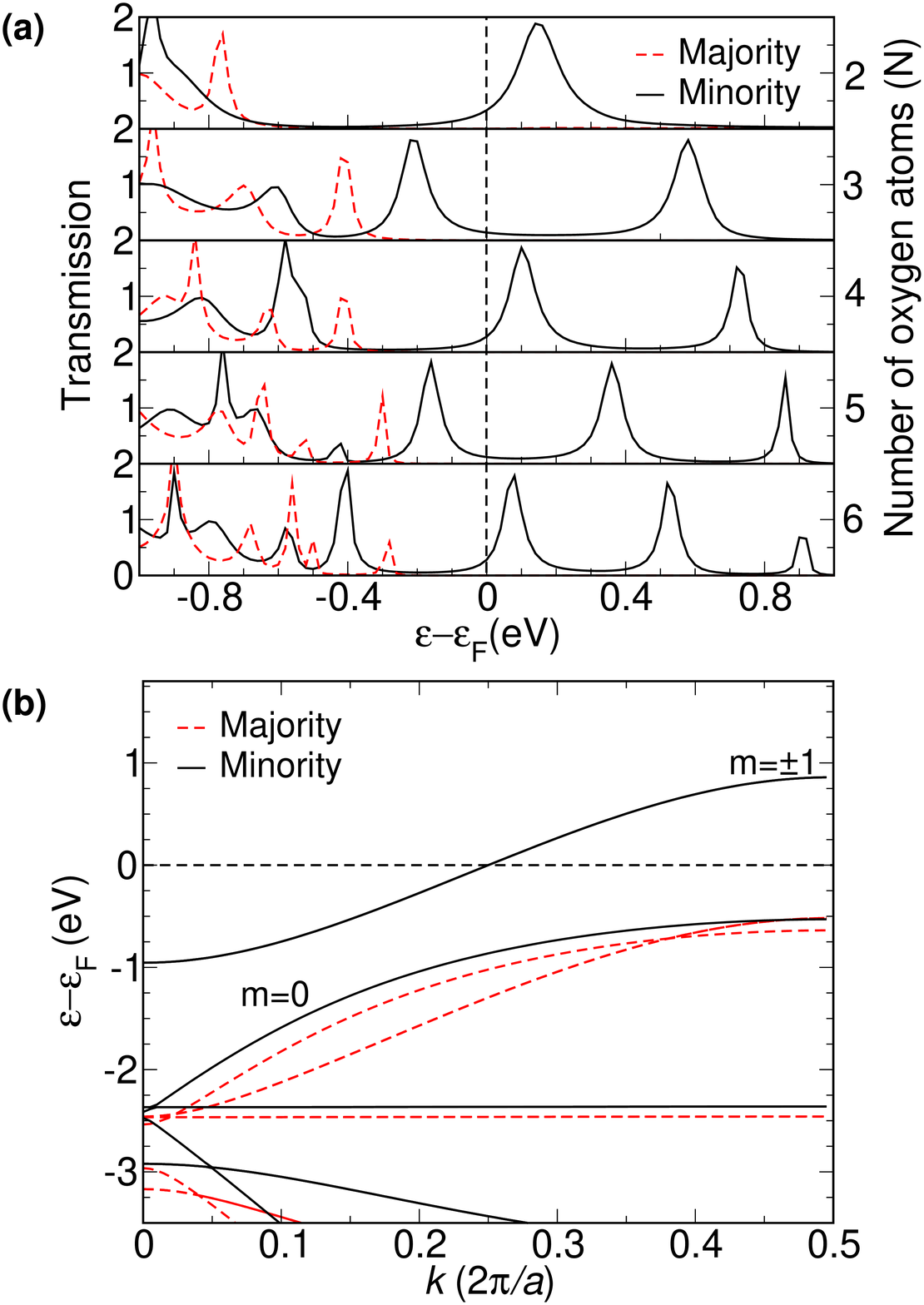}
   \caption{\label{fig.T_N_spin}(Color online) (a) Transmission
     functions for the majority spin (dashed line) and minority spin
     (full line) for Ag-O chains containing 2-6 oxygen atoms. (b) Band
     structure of the infinite Ag-O chain. The band crossing the Fermi
     level is doubly degenerate and the Bloch states have $m=\pm 1$
     symmetry with respect to rotations around the chain axis.}
\end{figure}

To gain more insight into the nature of the conductance oscillations,
we show in Fig.~\ref{fig.T_N_spin}a how the energy dependent
transmission functions change with $N$. The number of oxygen atoms in
the ASOC varies from $N=2$ in the top panel to $N=6$ in the bottom
panel. We first notice that only minority spin states are present at
the Fermi level which means that the current will be fully spin
polarized. The peaks in the transmission function coincide with
resonances of the chain which are broadened by coupling to the
contacts. The number of resonances increase linearly with $N$ and
their width decrease as $1/N$ due to a reduction in the overlap
between the levels on the chain and the states in the contact.  The
resonances have the same character as the Bloch states forming the
valence band of the infinite Ag-O chain whose band structure is shown
in Fig.~\ref{fig.T_N_spin}b. The spin polarized valence band is two
times degenerate with angular momentum $m=\pm 1$ and Ag($4d$)-O($2p$)
character.

Clearly, the conductance oscillations arise because the Fermi level
intersects the nearest resonance closer to its center for even $N$
than odd $N$. In contrast to the situation for homogeneous chains,
however, the transport is never ``on resonance'' but always takes
place via a resonance tail. In particular this cannot be explained by
charge neutrality. Indeed, from Fig.~\ref{fig.T_N_spin}a it can be
seen that the resonances are always almost completely empty or filled,
and since each resonance can accommodate two electrons (due to orbital
degeneracy), it seems that all chains accommodate an even number of
electrons. But each Ag-O unit contains 5 valence electrons, and local
charge neutrality would therefore imply a half-filled resonance for
every second Ag-O unit added to the chain. On the other hand, the
small-amplitude oscillations around the experimental saturation value
of $0.1G_0$ arise exactly because the Fermi level always intersect the
tail of a resonance.

In order to rationalize the conductance behavior of the ASOCs, as well
as that of other chain systems, we propose the following simple
resonating-chain model, which is exact for coherent transport above a
certain chain length. An electron approaching the chain from the left
lead will have a certain transmission probability $T_1$ for entering
into the chain. Inside the chain the electron can propagate back and
forth between the two (identical) contacts, so that every time the
electron impinges on one of the contacts it is reflected with a
probability $R_1= 1 - T_1$ and it furthermore picks up a phase shift
which we denote by $\phi_1$. We furthermore assume that the electron
in the infinite chain is characterized by a single Fermi Block wave
vector $k$. The interference between all the reflected waves then
leads to a total transmission probability $T$ through the chain system
of
\begin{equation}\label{eq.T_model}
  T = \frac{1}{1+4\frac{R_1}{(1-R_1)^2} \sin^2(k L + \phi_1)},
\end{equation}
where $L$ denotes the distance between the two contacts. This
expression is exact for coherent transmission in the limit where the
reflections at the two ends can be considered independent, {\em i.e.}
it is required that the potential in the central part of the chain is
unaffected by the two contacts. It should be noted that there is a
certain arbitrariness in the definition of the distance $L$ between
the two contacts, however, different choices for $L$ also gives
different results for the phase shift $\phi_1$ so that
Eq.~\ref{eq.T_model} still holds. We take the points of reflections to
be at the tip atoms of the contacts.

For a given reflection probability $R_1$ the total transmission is
seen to vary with the chain length between a maximum of $1$ and a
minimum of $((1-R_1)/(1+R_1))^2$. However, due to the phase shift and
the discrete nature of the length of the chain the transmission
usually oscillates over a more narrow region.

\begin{figure}[!h]
  \includegraphics[width=0.85\linewidth]{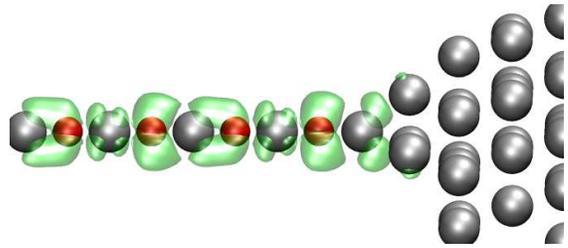}
  \caption{\label{fig.scattering_state} (Color online) Isosurface of
    the absolute value of the eigenchannel state $\Psi_s$ for a Ag-O
    chain. The state is a superposition of incoming and reflected
    chain Bloch states.}
\end{figure}

We have determined the parameters $R_1$ and $\phi_1$ for the ASOCs as
well as for chains of Al and Au. The parameters are obtained by
considering the reflection of an electron in a semi-infinite chain
impinging on a contact (see Fig.~\ref{fig.scattering_state}). We
determine the scattering state $|\Psi_s \rangle$ as described in
Ref.~\cite{magnus_scat}, and the phase shift is then determined by
projection onto the incoming ($k$) and outgoing ($-k$) states as
$\phi_1=\arg(\langle \psi_{-k}|\Psi_s\rangle /
\langle\psi_k|\Psi_s\rangle)$.

\begin{table}[hl]
  \begin{tabular}{|l|c|c|c|c|c|}
    \hline
    System & $R_1$ & $\phi_1$ & $k L$ & $T_{max}$ & $\Delta T_{osc}$\\
    \hline
    Ag-O/Ag  & 0.64 & $-0.21\pi$ & $N\pi/2$ & 0.12 & 0.05 \\
    Al/Al(111) & 0.32 & $0.92\pi$ & $(N+1)\pi/4$ &0.85 & 0.57\\
    Au/Au(100) & 0.004 & $-0.36\pi$ & $(N+1)\pi/2$ &0.997&0.010\\
    \hline
  \end{tabular}
  \caption{Calculated chain-contact reflection parameters for a number of chain systems.}
  \label{tab.scat}
\end{table}

In Table~\ref{tab.scat} the calculated reflection parameters for the
ASOCs and for mono-atomic Al and Au chains are shown together with the obtained
maximal transmissions and the magnitudes of the oscillations. In all
three cases the model reproduces the results of full DFT calculations
for varying chain lengths.

The model results for the ASOCs are shown in Fig.~\ref{fig.G_N} to
coincide with the full calculations for chains containing two or more
oxygen atoms. It should be noted that the low conductance of about
$~0.1 G_0$ comes about not so much because of the reduced reflection
probability ($R_1=0.64$) which would still allow for a total
transmission of 1 if a resonance condition could be met in the
chain. Rather it is the phase shift of $\phi_1=-0.21\pi \approx -
\pi/4$ which leads to destructive interference within the chain. The
actual size of this phase shift depends on both the electronic
structure of the chain and of the contacts and must be obtained from a
full calculation. However, we note that a phase shift of $\pi/4$
corresponds to a length of half a unit cell in the chain or
equivalently an oxygen-silver distance, and this is in good accordance
with Fig.~\ref{fig.scattering_state} where the scattering state is
seen to carry weight on the Ag tip atom and therefore is rather
reflected off the three-atom Ag layer below. We also mention that
Eq.~\ref{eq.T_model} reproduces the energy-dependent transmission
(Fig.~\ref{fig.T_N_spin}a) quite well including the resonant
structure, thus implicitly resolving the charge neutrality paradox --
odd numbers of chain electrons still doubly-occupy an integer number
of resonances when the effects of the phase shift at the boundaries
are included.

\begin{figure}[!h]
   \includegraphics[width=0.95\linewidth]{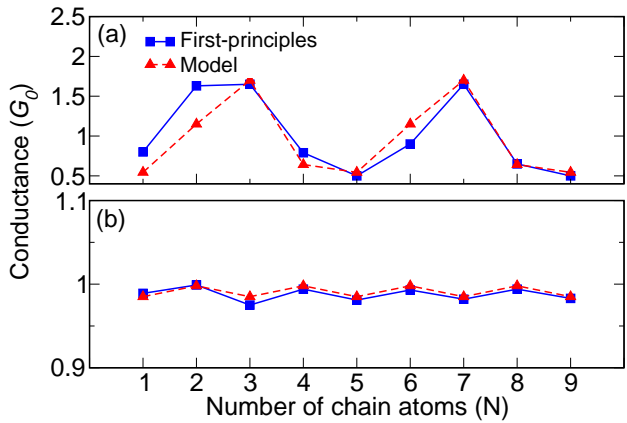}
   \caption{\label{fig.AlAu_G_N}(Color online) (a) Conductance for Al
     as function of the number of chain atoms, $N$. The
     first-principles result (squares) is taken from
     Ref.~\onlinecite{thygesen_alu} while model result (triangles) is
     obtained from Eq.~\ref{eq.T_model} with parameters derived using
     the same procedure as for the Ag-O system. (b) Conductance length
     dependence for Au with the first-principles result taken from
     Ref.~\onlinecite{brandbyge_osc}}.
\end{figure}

The results of the resonating-chain model for the conductances of Al
and Au chains are shown in Fig.~\ref{fig.AlAu_G_N} together with the
results of earlier full DFT-transport calculations. The agreement is
very good even for rather short chains. In the case of Al the phase
shift is close to $\pi$ in excellent agreement with the resonant-level
model proposed in Ref.~\cite{thygesen_alu}. The resonant-level model
takes as a starting point the isolated chain (of length $N$) which is
then coupled weakly to the leads, and this boundary condition
corresponds exactly to a phase shift of $\pi$. The small deviation of
the phase shift from $\pi$ accounts nicely for the fact that the
conductance does not peak at $2 G_0$ (the factor of two coming from
two available channels) but only at $1.7 G_0$.

For Au we find the phase and amplitude of the even-odd conductance
oscillations in agreement with the first-principles calculations in
Ref.~\onlinecite{brandbyge_osc}. We note, that the phase of the
conductance oscillation is not explained by either a charge neutrality
argument or a resonant model~\cite{thygesen_alu}.  However, the phase
of the oscillation can be related to our calculated phase shift of
$-0.36\pi$, while the small oscillation amplitude can be traced to an
almost perfectly transparent chain-bulk interface with a reflection
coefficient of $0.004$.

In summary, we have presented first principles conductance
calculations showing that alternating Ag-O chains suspended between
silver bulk contacts are half-metallic and can have an average
conductance of $0.1G_0$ as found in a recent
experiment~\cite{thijssen:026806}. In fact, the conductance oscillates
with a small amplitude and a period of two Ag-O units as the chain
length is varied. The oscillations can be understood from a
resonating-chain model, and are fully characterized by only two parameters:
The reflection probability, $R_1$, and reflection phase-shift,
$\phi_1$, of a single bulk-chain interface. By extracting these two
parameters from the DFT calculation, quantitative agreement between
the full calculations and the model is obtained.

We thank Jan van Ruitenbeek for illuminating discussions.  The authors
acknowledge support from the Danish Center for Scientific Computing
through grant HDW-1103-06. The Center for Atomic-scale Materials
Design is sponsored by the Lundbeck Foundation.

\bibliographystyle{apsrev}

\begin{thebibliography}{26}
\bibitem[{det({\natexlab{a}})}]{details}
\bibinfo{note}{We use an energy cutoff of $30$~Ry for the plane wave expansion
  and describe the ion cores by ultrasoft
  pseudopotensials~\cite{ultrasoft_psp}. For exchange and correlation we use
  the Perdew-Burke-Ernzerhof (PBE) functional~\cite{pbe_functional}. The
  Brillouin zone is sampled on a (1,4,4) Monkhorst pack grid.}
\bibitem[{det({\natexlab{b}})}]{details2} \bibinfo{note}{The
    SIESTA~\cite{SIESTA} calculations are based on normconserving
    pseudopotentials~\cite{Troullier_martin}, and a
    single-zeta-polarized basis for both Ag and O atoms. A mesh-cutoff
    of $200$~Ry is used for the electron density. We stress that the
    transmission function has been averaged over the Brillouin zone of
    the surface plane using a 4x4 $\mathbf{k_{\perp}}$-point
    sampling. This is crucial in order to avoid artificial features
    related to van-Hove singularities of the lead~\cite{kpoint}}.
\bibitem[{ban()}]{band} \bibinfo{note}{The Ag-O chain has two
    degenerate bands at the Fermi level, however, the corresponding
    transport channels are decoupled and can be treated separately.}
\end{thebibliography}

\end{document}